\documentclass[lettersize,journal]{IEEEtran}
\usepackage{amsmath,amsfonts}
\usepackage{setspace}
\usepackage{algorithm}
\usepackage{algpseudocode}
\usepackage{indentfirst}
\usepackage{xcolor}
\usepackage{array}
\usepackage[caption=false,font=normalsize,labelfont=sf,textfont=sf]{subfig}
\usepackage{textcomp}
\usepackage{stfloats}
\usepackage{url}
\usepackage{verbatim}
\usepackage{graphicx}
\def\BibTeX{{\rm B\kern-.05em{\sc i\kern-.025em b}\kern-.08em
    T\kern-.1667em\lower.7ex\hbox{E}\kern-.125emX}}
\usepackage{balance}
\usepackage{booktabs}
\usepackage{amssymb}
\usepackage[justification=centering]{caption}
\usepackage[justification=centering]{caption}
\usepackage{cite}
\newtheorem{theorem}{Theorem}
\newtheorem{lemma}{Lemma}
\newtheorem{proposition}{Proposition}
\newtheorem{remark}{Remark}
\begin{document}
\title{Rotatable Antenna Meets UAV: Towards Dual-Level Channel Reconfiguration Paradigm for ISAC}
\author{Shiying Chen, Guangji Chen, Long Shi, Qingqing Wu, and Kang Wei \vspace{-24pt}
\thanks{Shiying Chen, Guangji Chen, and Long Shi are with Nanjing University of Science and Technology, Nanjing 210094, China (email:\{shiyingchen, guangjichen, longshi\}@njust.edu.cn). Qingqing Wu is with Shanghai Jiao Tong University, 200240, China (e-mail: qingqingwu@sjtu.edu.cn). Kang Wei is with Southeast University, Nanjing, 211189, China (e-mail: kang.wei@seu.edu.cn)
}}


\maketitle
 \begin{abstract}
Integrated sensing and communication (ISAC) is viewed as a key enabler for future wireless networks by sharing the hardware and wireless resources between the functionalities of sensing and communication (S\&C). Due to the shared wireless resources for both S\&C, it is challenging to achieve a critical trade-off between these two integrated functionalities. To address this issue, this paper proposes a novel dual-level channel reconfiguration framework for ISAC by deploying rotatable antennas at an unmanned aerial vehicle (UAV), where both the large-scale path loss and the correlation of S\&C channels can be proactively controlled, thereby allowing a flexible trade-off between S\&C performance. To characterize the S\&C tradeoff, we aim to maximize the communication rate by jointly optimizing the RA rotation, the transmit beamforming, and the UAV trajectory, subject to the given requirement of sensing performance. For the typical scenario of static UAV deployment, we introduce the concept of subspace correlation coefficient to derive closed-form solutions for the optimal RA rotation, transmit beamforming, and UAV hovering location. For the scenario of a fully mobile UAV, we prove that the optimal trajectory of a UAV follows a hover–fly–hover (HFH) structure, thereby obtaining its global optimal solution. Simulation results show that the proposed design significantly improves the achievable S\&C trade-off region compared to benchmark schemes.
\end{abstract}
\vspace{-8pt}
\begin{IEEEkeywords}
ISAC, rotatable antenna, UAV.
\end{IEEEkeywords}

\section{Introduction}

Integrated sensing and communication (ISAC) is a crucial enabler for future wireless networks, offering the potential to deliver both ubiquitous connectivity and pervasive sensing in a spectrum and energy-efficient manner~\cite{ref1, ref2, ref3}. The shared exploitation of wireless resources, e.g., hardware platforms, wireless spectrum, and energy resources, significantly improves resource utilization efficiency, namely integration gain. To harvest the integration gain, conventional transceiver design should balance the transmit beam among S\&C channel subspaces, which leads to the performance loss when the S\&C channels are weakly coupled. Moreover, compared to the standalone communication network, the coverage issue of ISAC is more complicated due to the two-hop path loss for sensing services~\cite{ref4}. These limitations make it challenging to achieve optimal communication and sensing performance simultaneously.

To address the aforementioned challenges, the integration of unmanned aerial vehicles (UAVs) and rotatable antennas (RAs) has been regarded as a promising solution. In particular, UAVs have been viewed as a promising ISAC platform due to their three-dimensional (3D) mobility and flexible deployment to create line-of-sight (LoS) air-to-ground (A2G) links and to flexibly control the path loss of S\&C channels~\cite{ref5}. Beyond terrestrial ISAC, UAVs can act as aerial base stations (BSs), creating new opportunities for wide-area sensing coverage and dynamic communication enhancement\cite{ref6, ref7, ref8}. Regarding the issue of the weak coupling between S\&C channels in ISAC systems, a newly emerged flexible-antenna system, namely rotatable antennas (RAs), can leverage their dynamic orientation capability to improve channel coupling and boost beamforming gain~\cite{ref9,ref10}. By dynamically adjusting antenna orientation, RAs provide an additional spatial degree of freedom (DoF), enabling adaptive beam pattern shaping to improve channel alignment between communication and sensing links. While earlier studies have mainly focused on leveraging RAs to enhance the communication service~\cite{ref10, ref11}, the integration of RAs in ISAC systems is still in its infancy. 

Motivated by these advances, this paper proposes a novel RA-enabled UAV framework to facilitate the dual-level channel reconfiguration for ISAC, where the UAV mobility and the RA rotation are exploited for controlling the large-scale path loss and the small-scale coupling of S\&C channels, respectively, thereby achieving a more flexible trade-off between the performance of S\&C. To this end, we formulate an optimization problem to maximize the communication rate by jointly optimizing the UAV trajectory, RA rotation, and transmit beamforming, subject to the sensing performance requirement and UAV mobility constraints. For the static deployment scenario, we introduce the concept of correlation coefficient to determine the optimal RA rotation angle analytically, and then derive closed-form solutions for the transmit beamforming and UAV hovering location. For the fully mobile UAV scenario, we prove that the optimal trajectory follows a hover–fly–hover (HFH) structure. With the obtained insight, we perform a two-dimensional (2D) search over the optimal hovering location and accordingly obtain a closed-form expression of the optimal hovering time allocation to the original optimization problem. Numerical results corroborate that the proposed design significantly improves the achievable S\&C trade-off region compared with fixed antenna-based UAV ISAC schemes, in both static and fully mobile UAV scenarios.

\begin{figure}[!t]
\centering\includegraphics[width=3 in]{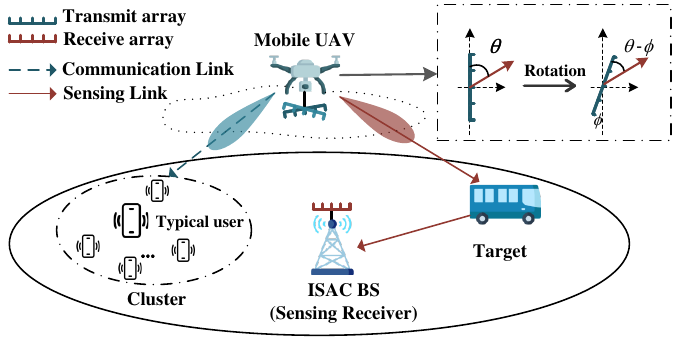}
\caption{The schematic of RA-enabled ISAC systems.}
\label{fig_1}
\vspace{-15pt} 
\end{figure}

\vspace{-8pt}
\section{System Model}
As illustrated in Fig.~1, we consider a bistatic ISAC system where a UAV equipped with a uniform linear RA-array of $M_t$ transmit antennas simultaneously senses one target on the ground while providing communication services for a cluster of single-antenna users. For ease of analysis, we focus on a typical user located in the center of the user cluster. The echo signals illuminated by the target are processed at the ground BS (GBS), equipped with a uniform linear array of $M_r$ receive antennas.
For convenience, we adopt a 2D Cartesian coordinate system, where the typical user and target are located at $\mathbf{q}_{\mathrm u} = [0,0]^T$ and $\mathbf{q}_{\mathrm t} = [D_{\mathrm x}, D_{\mathrm y}]^T$, respectively. The UAV flies at a constant altitude $H$ with flight duration $T$, and its horizontal position at time $t \in \mathcal{T} = [0,T]$ is denoted by $\mathbf{q}(t) = [x_q(t), y_q(t)]^T$. All A2G channels are assumed to be dominated by LoS components~\cite{ref6}. The wireless channel from the UAV to the user is given by
\begin{equation}
{{\bf{h}}_{\rm{c}}}({\bf{q}}(t),\phi (t)) = \sqrt {{\beta _0}} {\mkern 1mu} \frac{{{e^{ - {\rm{j}}{\mkern 1mu} {\textstyle{{2\pi } \over \lambda }}{\mkern 1mu} {d_{\rm{c}}}({\bf{q}}(t))}}}}{{{d_{\rm{c}}}({\bf{q}}(t))}}{\mkern 1mu} {{\bf{a}}_{\rm{t}}}({\theta _{\rm{u}}}({\bf{q}}(t)) - \phi (t)).
\end{equation}
where $\beta_0$ represents the channel power gain at the reference distance of 1 meter (m), $\lambda$ represents the carrier wavelength, $d_c(\mathbf{q}(t)) = \sqrt{\|\mathbf{q}(t) - \mathbf{q}_u\|^2 + H^2}$ is the distance from the UAV to the user, $\theta_u(\mathbf{q}(t))$ denotes the spatial angle between UAV and user, and $\phi(t)$ is the rotation angle of the antenna array. The transmit steering vector is given by $\mathbf{a}_t(\theta) = [1,\, e^{j\pi\sin\theta},\, \ldots,\, e^{j\pi(M_t-1)\sin\theta}]^T$ with $\theta$ denoting the azimuth angle of departure (AoD) from the UAV to the user.
 At time $t$, the ISAC transmit signal is $\mathbf{x}(t) = \mathbf{w}(t)\, s_c(t)$, where $\mathbf{w}(t) \in \mathbb{C}^{M_t \times 1}$ is the transmit beamforming vector and $s_c(t) \sim \mathcal{CN}(0,1)$ is the communication symbol. Thus, the received signal at the user is $y(t)=\mathbf h_{\mathrm c}^{\mathrm H}\!\big(\mathbf q(t),\phi(t)\big)\,\mathbf x(t)+n_{\mathrm c}(t)
$, where $n_c(t) \sim \mathcal{CN}(0, \sigma_c^2)$ denotes the additive white Gaussian noise (AWGN) at the user. Accordingly, the achievable rate is
\begin{align}
    R(\mathbf q(t),\phi(t),\mathbf w(t)) &= \log_2\!\Bigl(1+\gamma_{\mathrm c}(\mathbf q(t),\phi(t),\mathbf w(t))\Bigr),
\end{align}
where $\gamma_{\mathrm c}(\mathbf q(t),\phi(t),\mathbf w(t)) = {\bigl|\mathbf h_{\mathrm c}^{\mathrm H}\!\bigl(\mathbf q(t),\phi(t)\bigr)\,\mathbf w(t)\bigr|^{2}}/{\sigma_{\mathrm c}^{2}}$.

For the wireless sensing service, the wireless channel of the UAV--target--GBS link is modeled as
\begin{equation}\label{eq:Gs_mr}
\mathbf G_{\mathrm s}\bigl(\mathbf q(t),\phi(t)\bigr)
= \alpha\,\frac{\sqrt{\beta_0}}{d_s(\mathbf q(t))}\,
\mathbf a_{\mathrm r}(\theta_{\mathrm r})\,
\mathbf a_{\mathrm t}^{\mathrm H}\!\bigl(\theta_{\mathrm s}(\mathbf q(t))-\phi(t)\bigr).
\end{equation}
where $\alpha$ is the complex reflection coefficient of the target, $d_s(\mathbf{q}(t)) = \sqrt{\|\mathbf{q}(t) - \mathbf{q}_t\|^2 + H^2}$ is the distance from the UAV to the target, $\theta_s(\mathbf{q}(t))$ denotes the spatial angle from the UAV to the target relative to the UAV’s array orientation, and $\theta_r$ denotes the angle from the target to the GBS relative to the array orientation of the GBS. The UAV’s transmit steering vector $\mathbf{a}_t(\cdot)$ accounts for the RA rotation $\phi$, while the steering vector of the GBS $\mathbf{a}_r(\cdot)$ depends only on $\theta_r$. During the transmission period, the received echo signal at the GBS is given by $\mathbf{y}_r(t) = \mathbf{G}_s(\mathbf{q}(t),\phi(t))\mathbf{x}(t) + \mathbf{n}_r(t)$, where $\mathbf{n}_r(t)\sim\mathcal{CN}(0, \sigma_r^2 \mathbf{I})$ denotes the additive noise at the GBS. 
It is well recognized that the performance of the general sensing tasks is closely related to the illuminated signal power from the target. Intuitively, a higher signal power illuminating the target results in stronger echo signals at the GBS, thereby enhancing detection probability or estimation accuracy. Without loss of generality, we adopt the sensing signal-to-noise ratio (SNR) as a unified metric to evaluate the sensing performance. Accordingly, the sensing SNR with the UAV position $\mathbf{q}(t)$ and the rotation angle $\phi(t)$ is
\begin{equation} 
\Gamma_s(\mathbf q(t), \phi(t), \mathbf w(t)) = {\left\| \mathbf G_s\big(\mathbf q(t), \phi(t)\big)\, \mathbf w(t) \right\|^2}/{\sigma_r^2}. 
\end{equation}
where $\sigma_r^2$ is the noise power at the GBS. Here, $\|\mathbf a_r(\theta)\|^2 = M_r$.

To characterize the fundamental trade-off between S\&C, we aim to maximize the achievable communication rate by jointly optimizing the UAV trajectory~$\mathbf{q}(t)$, RA rotation~$\phi(t)$, and beamforming~$\mathbf{w}(t)$, subject to sensing SNR, UAV flight speed, and the maximum transmit power constraint. The corresponding optimization problem is formulated as
\begin{subequations}\label{eq:P1}
\begin{align}
    \max_{\mathbf{q}(t),\,\phi(t),\,\mathbf{w}(t)} \quad
    & \frac{1}{T}\int_0^T R\big(\mathbf{q}(t),\phi(t),\mathbf{w}(t)\big)  \, dt \label{P1_obj}\\
    \mathrm{s.t.}\quad
    &\frac{1}{T}\int_0^T \Gamma_s\big(\mathbf{q}(t),\phi(t),\mathbf{w}(t)\big) \, dt \geq \Gamma_\mathrm{th}, \label{P1_sensing}\\
    & \|\mathbf{w}(t)\|^2 \leq P_{\max}, \quad \forall t\in[0,T] \label{P1_power}, \\ 
    & \|\dot{\mathbf{q}}(t)\| \leq V, \quad \forall t\in[0,T] \label{P1_speed},
\end{align}
\end{subequations}
where constraint~(\ref{P1_sensing}) guarantees that the average sensing SNR satisfies the predefined threshold $\Gamma_{\mathrm{th}}$, ~(\ref{P1_power}) enforces the transmit power budget, and (\ref{P1_speed}) restricts the UAV flight velocity, with $V$ being the UAV's maximum horizontal flight speed. Here, $\dot{\mathbf q}(t) = \frac{d\mathbf q(t)}{dt}$ denotes its horizontal velocity vector. It is challenging to obtain the optimal solution via standard methods due to the following reasons. First, problem (\ref{eq:P1}) is non-convex due to the non-concave objective function and coupled optimization variables in constraint (\ref{P1_sensing}). Second,  problem (\ref{eq:P1}) involves an infinite number of variables, i.e., $\mathbf{q}(t)$, $\phi(t)$, and $\mathbf{w}(t)$. Nevertheless, we derive its optimal solution by exploiting its particular structure in the next section.

\section{Proposed Solution}
We first optimize the antenna rotation angle and transmit beamforming vector of problem (5) under the given $\mathbf q(t)$. Let $\mathbf h_c = \mathbf h_c\!\big(\mathbf q(t), \phi(t)\big)$, 
$\mathbf h_t = \mathbf a_t\!\big(\theta_s(\mathbf q(t)) - \phi(t)\big)$, and
$\tilde{\Gamma}=\Gamma_{\mathrm{th}}\sigma_r^2/\alpha^2$. For any given $(\mathbf{q}(t),\phi(t))$, the maximum communication SNR admits the following closed-form expression.

\begin{lemma} 
With the optimal beamforming vector, the maximum communication SNR is
\begin{equation}
\mathrm{SNR}_c^*\!\big(\mathbf{q}(t),\phi(t)\big)=
\begin{cases}
\dfrac{P_{\max}\,\|\mathbf h_c\|^2}{\sigma_c^2},\quad
\text{if }\, P_{\max}\,\rho^2\,\|\mathbf h_t\|^2 \ge \tilde{\Gamma}, \\
g(\rho),\quad \text{Otherwise},
\end{cases}
\end{equation}
where $g(\rho) =\frac{\|\mathbf h_c\|^2}{\sigma_c^2\,\|\mathbf h_t\|^2}\Big(\rho\sqrt{\tilde{\Gamma}}+\sqrt{1-\rho^2}\,\sqrt{\|\mathbf h_t\|^2 P_{\max}-\tilde{\Gamma}}\Big)^{\!2}$, and the correlation coefficient is $\rho = \frac{|\mathbf h_c^H\mathbf h_t|}{\|\mathbf h_c\|\,\|\mathbf h_t\|}\in[0,1]$. 
\end{lemma} 

\textit{Proof:} See~\cite{ref12}. \hfill$\blacksquare$

Lemma 1 implies that the optimal rotation angle $\phi^*(t)$ should be designed to maximize $\rho$ for enhancing the coupling of S\&C channels. To this end, we have the following theorem to provide closed-form expressions of $\phi^*(t)$ and $\mathbf{w}^*(\mathbf{q}(t))$.

\begin{theorem}
For $\forall t$, the optimal rotation angle and beamforming vector of problem (\ref{eq:P1}), denoted by $\phi^*(t)$ and $\mathbf w^*\!\big(\mathbf q(t)\big)$, are given by 
\begin{equation}
    \phi^*(t) = \frac{\theta_u(\mathbf{q}(t))+\theta_s(\mathbf{q}(t))}{2} \pm \frac{\pi}{2}, \label{eq:opt_bf}
\end{equation}
\begin{equation}
\mathbf w^*\!\big(\mathbf q(t)\big)
= \sqrt{P_{\max}}\,
\frac{\mathbf h_c\!\big(\mathbf q(t),\,\phi^*(t)\big)}
     {\big\|\mathbf h_c\!\big(\mathbf q(t),\,\phi^*(t)\big)\big\|},
\end{equation}
respectively. The corresponding SNRs of S\&C are 
\begin{equation}
\mathrm{SNR}_c(\mathbf{q}(t))=\frac{P_{\max}\beta_0 M_t}{\sigma_c^2\,d_c^2(\mathbf{q}(t))}, 
\mathrm{SNR}_s\bigl(\mathbf q(t)\bigr) = \frac{P_{\max}\,|\alpha|^2\,\beta_0\,M_t M_r}{\sigma_r^2\,d_s^2\!\bigl(\mathbf q(t)\bigr)}.
\end{equation}
\end{theorem}

\textit{Proof:} Let $\zeta_1(t) = \frac{\theta_u(\mathbf{q}(t))-\theta_s(\mathbf{q}(t))}{2}$ and $\zeta_2(t) = \frac{\theta_u(\mathbf{q}(t))+\theta_s(\mathbf{q}(t))}{2}$. Then, we have a compact form as
\begin{equation}
    \rho(t,\phi) = \left|\frac{\sin\left(M_t\pi\sin\zeta_1(t)\cos(\zeta_2(t)-\phi)\right)}{M_t\sin\left(\pi\sin\zeta_1(t)\cos(\zeta_2(t)-\phi)\right)}\right|.
\end{equation}
It is evident that $\big|\sin(M_t\pi x)/(M_t\sin\pi x)\big|\le 1$ where the equality holds if and only if $x\in\mathbb Z$. Define
$x = \sin \zeta_1(t)\,\cos\bigl(\zeta_2(t)-\phi\bigr)$ and then we have $x \in \left[ { - 1,1} \right]$ since $\bigl|\sin\zeta_1(t)\,\cos\bigl(\zeta_2(t)-\phi\bigr)\bigr| \le 1$. Hence, $\max_\phi \rho(t,\phi)=1$ holds if and only if $\sin\zeta_1(t)\cos(\zeta_2(t)-\phi)=0$, which leads to $\phi^*(t)$ in (7). At $\phi=\phi^*(t)$, maximizing $|\mathbf h_c^H\mathbf w|$ subject to the power and sensing constraints reduces to align $\mathbf w$ with $\mathbf h_c$, which yields the maximum ratio transmission (MRT) solution in (8).
\hfill$\blacksquare$

Using (7)–(9) in Theorem 1, the instantaneous rate of the UE and the sensing SNR at the GBS are
\begin{align}
R\big(\mathbf q(t)\big)
  &= \log_2\!\left(1+\mathrm{SNR}_c\!\big(\mathbf q(t)\big)\right),\\
\Gamma\!\big(\mathbf q(t)\big)
  &= \frac{\alpha^2 P_{\max}\beta_0\, M_t}
          {\sigma_r^2\!\left(\big\|\mathbf q(t)-\mathbf q_t\big\|^2+H^2\right)}.
\end{align}

Theorem~1 indicates that exploiting the rotation of RAs can perfectly align the subspaces of S\&C channels, thereby leading to simple MRT beamforming that simultaneously maximizes the channel power gain for both links. Theorem~1 lays the foundation for the subsequent optimization of the UAV trajectory. Inspired by Theorem~1, problem (\ref{eq:P1}) is simplified as 
\begin{subequations}\label{eq:P2}
\begin{align}
\max_{\mathbf{q}(t)}\;\;& \frac{1}{T}\int_{0}^{T} R\big(\mathbf{q}(t)\big)\,dt \\[2pt]
\text{s.t.}\;\;& \frac{1}{T}\int_{0}^{T} \Gamma\big(\mathbf{q}(t)\big)\,dt \;\ge\; \Gamma_{\mathrm{th}}, \\[2pt]
& \|\dot{\mathbf{q}}(t)\| \;\le\; V,\quad t\in[0,T].
\end{align}
\end{subequations}
\vspace{-15pt} 

\subsection{Static UAV Deployment}
Problem~\eqref{eq:P2} is still challenging to be solved since it contains the non-concave objective function and an infinite number of variables. To overcome this issue, we first focus on a special case of $V \to 0$ in this subsection, which corresponds to a typical scenario of UAV deployment optimization where the UAV hovers at a fixed location once deployed. 
\subsubsection{Proposed Solution}
For ease of exposition, we drop the index of time $t$, and problem (13) is given by
\begin{subequations}\label{eq:P3}
\begin{align}
    \max_{\mathbf{q}} \quad & R(\mathbf{q}) \\
    \text{s.t.} \quad & \Gamma(\mathbf{q}) \geq \Gamma_{\text{th}}.
\end{align}
\end{subequations}
The maximum sensing SNR is achieved by placing the UAV at $\mathbf q_t$, which leads to $\Gamma_{\max}=\max_{\mathbf q}\,\Gamma(\mathbf q) =\frac{\alpha^2 P_{\max}\beta_0 M_t}{\sigma_r^2 H^2}$. Hence, problem \eqref{eq:P3} is feasible provided that
\begin{equation}\label{eq:fea_reg}
\Gamma_{\mathrm{th}}\ \le\ \frac{\alpha^2 P_{\max}\beta_0 M_t}{\sigma_r^2 H^2}.
\end{equation}
Given the feasibility condition in \eqref{eq:fea_reg}, we now derive the optimal solution to problem~\eqref{eq:P3} in the following theorem.

\begin{theorem} 
The optimal solution to problem~\eqref{eq:P3}, denoted by ${{\mathbf{q}}^*}$, is given by
\begin{equation}\label{eq:qstar}
    \mathbf q^* = 
    \begin{cases}
        \mathbf q_u, & D_h \leq R_h, \\
        \mathbf q_t + \dfrac{R_h}{D_h}(\mathbf q_u - \mathbf q_t), & D_h > R_h,
    \end{cases}
\end{equation}
where $D_h=\|\mathbf q_u-\mathbf q_t\|$, $R_s= \sqrt{\frac{\alpha^2 P_{\max}\beta_0 M_t}{\Gamma_{\mathrm{th}}\sigma_r^2}}$, $R_h= \sqrt{R_s^2-H^2}$. With $\mathbf q^*$ of (16), the corresponding optimal objective value of problem~\eqref{eq:P3} is
\begin{equation}\label{eq:Rstar}
R^*(\Gamma_{\mathrm{th}})=
\begin{cases}
\displaystyle \log_2\!\left(1+\frac{A}{H^2}\right),
& 0\le \Gamma_{\mathrm{th}}\le \dfrac{B}{D_h^2+H^2},\\[8pt]
\displaystyle f(\Gamma_{\mathrm{th}}),
& \dfrac{B}{D_h^2+H^2}<\Gamma_{\mathrm{th}}\le \dfrac{B}{H^2},\\[8pt]
\end{cases}
\end{equation}
where $f(\Gamma_{\mathrm{th}})=\log_2\!\left(1+\frac{A}{\left(D_h-\sqrt{\frac{B}{\Gamma_{\mathrm{th}}}-H^2}\right)^{2}+H^2}\right)$, $A=\frac{P_{\max}\beta_0 M_t}{\sigma_c^2}$ and $B=\frac{\alpha^2 P_{\max}\beta_0 M_t}{\sigma_r^2}$.
\end{theorem}

\textit{Proof:} Please refer to Appendix A.\hfill$\blacksquare$

The achievable region of the communication rate and sensing SNR in the static deployment case is defined as ${{\cal C}_{{\mathop{\rm sta}\nolimits} }}$. Using \eqref{eq:qstar} and \eqref{eq:Rstar} in Theorem 2, we unveil an essential property of ${{\cal C}_{{\mathop{\rm sta}\nolimits} }}$ in the following proposition.

\begin{proposition}
 Under the static UAV deployment, the achievable region of the communication rate and the sensing SNR is
\begin{equation}
\mathcal C_{\rm sta} = \Big\{(\Gamma_{\mathrm{th}},R):\ 0\le\Gamma_{\mathrm{th}}\le \tfrac{B}{H^2},\ R\le R^*(\Gamma_{\mathrm{th}})\Big\},
\end{equation}
where ${{\cal C}_{{\mathop{\rm sta}\nolimits} }}$ is non-convex if $D_h^2 > \tfrac{4}{3}\sqrt{H^2 + A}\bigl(\sqrt{4H^2 + A}-\sqrt{H^2 + A}\bigr)$.
\end{proposition}

\textit{Proof:} Please refer to Appendix B.
\hfill$\blacksquare$\par\noindent
Proposition 1 implies that whenever the user–target horizontal separation $D_h$ exceeds a certain threshold, the rate–sensing region $\mathcal C_{\rm sta}$ is non-convex.
\begin{remark}
For the non-convex static region $\mathcal C_{\rm sta}$, it is evident that $\mathcal C_{\rm sta}\subseteq\operatorname{Conv}(\mathcal C_{\rm sta})$, where $\operatorname{Conv}(\mathcal C_{\rm sta})$ is the convex hull operation of $\mathcal C_{\rm sta}$, i.e.,
$\operatorname{Conv}(\mathcal C_{\rm sta})=\{\sum_{n}\alpha_n z_n:\ z_n\in\mathcal C_{\rm sta},\ \alpha_n\ge0,\ \sum_{n}\alpha_n=1\}$. It demonstrates that any point on the boundary of $\operatorname{Conv}(\mathcal C_{\rm sta})$ can be obtained by performing time sharing among the two points on the boundary of ${{\cal C}_{{\mathop{\rm sta}\nolimits} }}$, which unveils that exploiting the UAV mobility can further enlarge the region of ${{\cal C}_{{\mathop{\rm sta}\nolimits} }}$. The details will be elaborated in the following numerical results.
\end{remark}

\subsubsection{Numerical Results}

\begin{figure}[!t]
\centering\includegraphics[width=3.3 in]{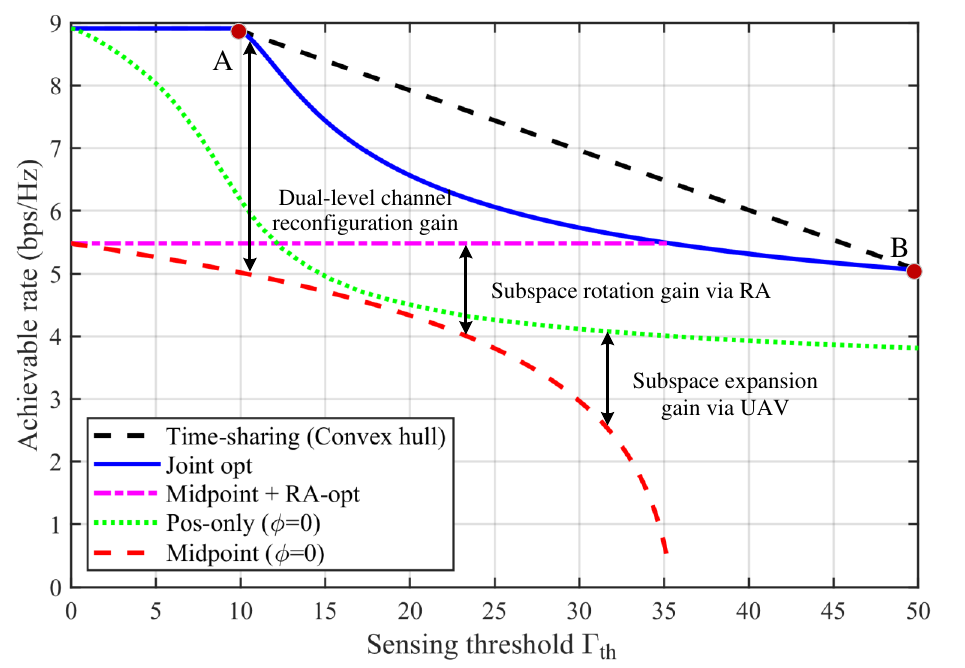}
\caption{Achievable rate versus sensing threshold under the static UAV deployment.}
\label{fig_2}
\vspace{-15pt} 
\end{figure}

The simulation parameters are set as follows: $M_t=12$, $M_r=16$, $P_{\max}=0.1$\,W, $\beta_0=10^{-4}$, $\sigma_c^2=10^{-18}$, $\sigma_r^2=10^{-18}$, $\alpha=0.9$, $\lambda=0.03$\,m, $H=50$\,m, $\mathbf{q}_u=[0,0]$, $\mathbf{q}_t=[300,100]$. 
We compare the following schemes: 
1) \textbf{Midpoint} ($\phi\!=\!0$): the UAV is fixed at $[150,50]\,\mathrm{m}$ and the array orientation is fixed; 
2) \textbf{Pos-only} ($\phi\!=\!0$): the UAV position is optimized along the user–target line while the array orientation remains fixed; 
3) \textbf{Midpoint+RA-opt}: the UAV is fixed at $[150,50]\,\mathrm{m}$ while the RA rotation angle is optimized; 
4) \textbf{Joint opt}: the UAV hovering location and RA rotation are jointly optimized; and 
5) \textbf{TS-bound}: the convex hull operation of the \textbf{Joint opt} boundary, serving as the performance upper bound of the performance under the fully mobile UAV.

Fig.~2 illustrates the achievable communication rate versus the sensing SNR threshold under all the considered schemes. It is observed that the proposed \textbf{Joint opt} consistently yields the largest achievable rate over the entire range of sensing thresholds compared with the benchmarks, which underscores the importance of jointly exploiting the rotation of RAs and UAV position optimization for improving both the coupling and strength of S\&C channels. Moreover, any point on the black dashed line can be obtained by performing time-sharing among the two transmit schemes corresponding to the points A and B, respectively. It is observed that the rates obtained on the boundary of \textbf{TS-bound} are superior to those achieved by the static UAV deployment, thereby highlighting that exploiting the mobility of the UAV can further enlarge the region of the achievable S\&C performances.

\subsection{Fully Mobile UAV Case}
The analytical results in the previous subsection imply that the achievable region of S\&C performance is further improved by exploiting the UAV mobility. In this subsection, we focus on the general case of dynamic trajectory optimization of a UAV.
\subsubsection{Proposed Solution}
To draw useful insights, we unveil the structure of the optimal UAV trajectory under the special case of $V \to \infty $. The results are summarized in the following remark.

\begin{remark}
When $V\to\infty$, the flight time between any two points is negligible. In this case, the continuous trajectory reduces to time-sharing among hovering locations, and the achievable region of S\&C performance is the convex hull of the Pareto boundary obtained under the case of the static UAV deployment. Hence, any point located at the boundary of the achievable region is realized by time-sharing between at most two static UAV hovering locations. This phenomenon has also been demonstrated in the black dashed line of Fig.~2. The result implies that the optimal UAV trajectory under the case of $V \to \infty$ is placing the UAV at two optimal hovering locations and allocating mission time between them. In particular, the maximum average rate is attained by the hovering locations $\mathbf q_a,\mathbf q_b $ that lie on the user–target line, with time allocations
\begin{equation}
\tau_a = \frac{T\big(\Gamma_{\rm th}-\Gamma(\mathbf q_b)\big)}{\Gamma(\mathbf q_a)-\Gamma(\mathbf q_b)},\qquad
\tau_b=T-\tau_a,
\end{equation}
where we assume $\Gamma(\mathbf q_a)>\Gamma(\mathbf q_b)$ for feasibility. The resultant maximum average rate is
\begin{equation}
\bar R^*=\max_{\mathbf q_a,\mathbf q_b}\frac{1}{T}\big(\tau_a R(\mathbf q_a)+\tau_b R(\mathbf q_b)\big).
\end{equation}
\end{remark}

Motivated by this observation, we construct an HFH trajectory. Specifically, the UAV first hovers at the initial location $\mathbf q_a $ for time $\tau_a$. Then, the UAV flies to the final location $\mathbf q_b$ at the maximum speed $V$ with the flight time $\Delta t_{\rm fly}=\|\mathbf q_a-\mathbf q_b\|/V$. Finally, the UAV hovers at the final location $\mathbf q_b $ for the time $\tau_b$.
Mathematically, the HFH trajectory is expressed as
\begin{equation}
\mathbf{q}(t) = 
\begin{cases} 
\mathbf{q}_a, & t \in [0, \tau_a) \\
\mathbf{q}_a + \frac{V(t - \tau_a)}{\|\mathbf{q}_b - \mathbf{q}_a\|} (\mathbf{q}_b - \mathbf{q}_a), & t \in [\tau_a, \tau_a + \Delta t_{\text{fly}}) \\
\mathbf{q}_b, & t \in [\tau_a + \Delta t_{\text{fly}}, T]
\end{cases}
\end{equation}
with total mission time satisfying \( \tau_a + \tau_b + \Delta t_{\text{fly}} = T \). Then, we unveil the optimality of the HFH structure for the general case of finite \( V \) through the following theorem.

\begin{theorem}
For any finite speed $V>0$, the HFH trajectory is optimal to problem (13). 
\end{theorem}

\begin{figure}[!t]
\centering\includegraphics[width=2.2 in]{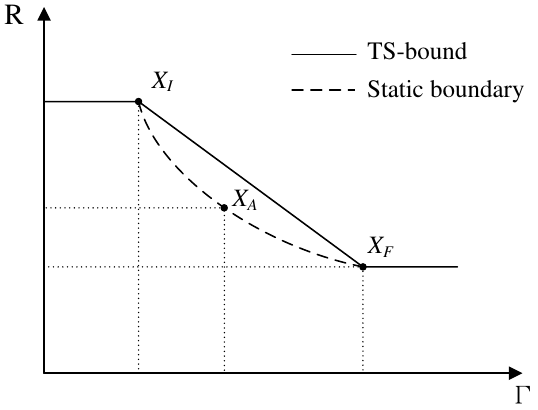}
\caption{ Illustration of the TS-bound and the static boundary induced by \(C_f(x_I)\) and \(C_f(x_F)\).}
\label{fig_3}
\vspace{-15pt} 
\end{figure}

\textit{Proof:}
Let $x_I$ and $x_F$ be the two endpoints on the user--target line that maximize communication and sensing. Define $z(x) = \big(\Gamma(x), R(x)\big)$, and let $C_f(x)$ denote the achievable rate–sensing region when the UAV hovers at location $x$. As illustrated in Fig.~\ref{fig_3}, the static boundary $z(x)$ on $[x_I,x_F]$ lies below the time-sharing bound determined by $z(x_I)$ and $z(x_F)$. Hence for any interior $x_A\in(x_I,x_F)$ there exists $\mu\in(0,1)$ such that
\begin{equation}
z(x_A)\leq \ \mu\,z(x_I)+(1-\mu)\,z(x_F),
\end{equation}
with at least one strict inequality unless $x_A\in\{x_I,x_F\}$. Therefore, $C_f(x_A)\subseteq \operatorname{Conv}\big(C_f(x_I)\cup C_f(x_F)\big)$, which implies that any achievable performance at $x_A$ is equivalently obtained by proper time-sharing between $x_I$ and $x_F$. Hence, hovering at $x_A$ is strictly redundant. Therefore, the UAV should only hover at the boundary points $x_I$ and $x_F$, while flying between them at maximum speed $V$. \hfill$\blacksquare$

With the constructed HFH trajectory in (21), problem (13) is reduced to the joint optimization of \( (\mathbf{q}_a, \mathbf{q}_b) \) and time allocation \( (\tau_a, \tau_b) \), which greatly simplify the algorithm design by reducing the number of optimization variables. To this end, we define the total hovering time as $T_{\mathrm{hov}} = T - \Delta t_{\mathrm{fly}}$, and introduce a time-sharing factor $\mu \in [0,1]$ such that $\tau_a = \mu T_{\mathrm{hov}}$ and $\tau_b = (1-\mu) T_{\mathrm{hov}}$.
During the flying phase, the UAV's instantaneous position at time $u \in [0, \Delta t_{\mathrm{fly}}]$ is expressed as $\mathbf{q}_{\mathrm{fly}}(u) = \mathbf{q}_a + \frac{u}{\Delta t_{\mathrm{fly}}} (\mathbf{q}_b - \mathbf{q}_a)$. With these definitions, the average communication rate and the average sensing SNR, denoted by $\bar{R}$ and $\bar{\Gamma}$, are expressed as
\begin{align}
\bar{R} &= \frac{T_{\mathrm{hov}}}{T} \left[ \mu R(\mathbf{q}_a) + (1-\mu) R(\mathbf{q}_b) \right] + \frac{1}{T} \int_0^{\Delta t_{\mathrm{fly}}} R\big(\mathbf{q}_{\mathrm{fly}}(u)\big) du, \\
 \bar{\Gamma} &= \frac{T_{\mathrm{hov}}}{T} \left[ \mu \Gamma(\mathbf{q}_a) + (1-\mu) \Gamma(\mathbf{q}_b) \right] + \frac{1}{T} \int_0^{\Delta t_{\mathrm{fly}}} \Gamma\big(\mathbf{q}_{\mathrm{fly}}(u)\big) du, 
\end{align}

Accordingly, the equivalent optimization problem with respect to the hovering locations and time allocation is given by
\vspace{-0.6\baselineskip}
\begin{subequations}\label{eq:P4}
\begin{align}
\max_{\mathbf{q}_a, \mathbf{q}_b, \mu} \quad & \bar{R} \\
\text{s.t.} \quad & \bar{\Gamma} \geq \Gamma_{\mathrm{th}}, \\
& T_{\mathrm{hov}} \geq 0, \quad \mu \in [0, 1].
\end{align}
\end{subequations}
Problem \eqref{eq:P4} is solved optimally by a 2D search over two hovering locations satisfying $\mathbf{q}_a = \mathbf{q}_u + s_a (\mathbf{q}_t - \mathbf{q}_u)$, $\mathbf{q}_b = \mathbf{q}_u + s_b (\mathbf{q}_t - \mathbf{q}_u)$, $0 \le s_a \le s_b \le 1 $.
For each $(s_a,s_b)$, define the straight-line flight and available hovering time as
$\Delta t_{\rm fly}=\frac{(s_b-s_a)\,\|\mathbf{q}_t-\mathbf{q}_u\|}V$,
and the flight path is given by
\begin{equation}
\mathbf{q}_{\rm fly}(u)=\mathbf{q}_u + \Big[s_a + \frac{u}{\Delta t_{\rm fly}}(s_b-s_a)\Big](\mathbf{q}_t-\mathbf{q}_u).
\end{equation}
Let $I_\Gamma(\mathbf{q}_a,\mathbf{q}_b)=\int_{0}^{\Delta t_{\rm fly}}\Gamma\!\big(\mathbf{q}_{\rm fly}(u)\big)\,du$
be the cumulative sensing during flight. With hovering times, activating the average sensing constraint yields
\begin{equation}
\mu^*=\frac{\Gamma_{\rm th}T - I_\Gamma(\mathbf{q}_a,\mathbf{q}_b) - T_{\rm hov}\Gamma(\mathbf{q}_b)}
{T_{\rm hov}\big(\Gamma(\mathbf{q}_a)-\Gamma(\mathbf{q}_b)\big)} .
\end{equation}
Project $\mu^*$ onto $[0,1]$ to ensure feasibility, then set $\tau_a^*=\mu^* T_{\rm hov}, \tau_b^*=(1-\mu^*)T_{\rm hov}$.
Thus, the overall optimization reduces to a search over $(s_a,s_b)$, with a closed-form time allocation $\mu^*$ evaluated for each candidate.

\subsubsection{Numerical Results}

\begin{figure}[!t]
\centering\includegraphics[width=3.3 in]{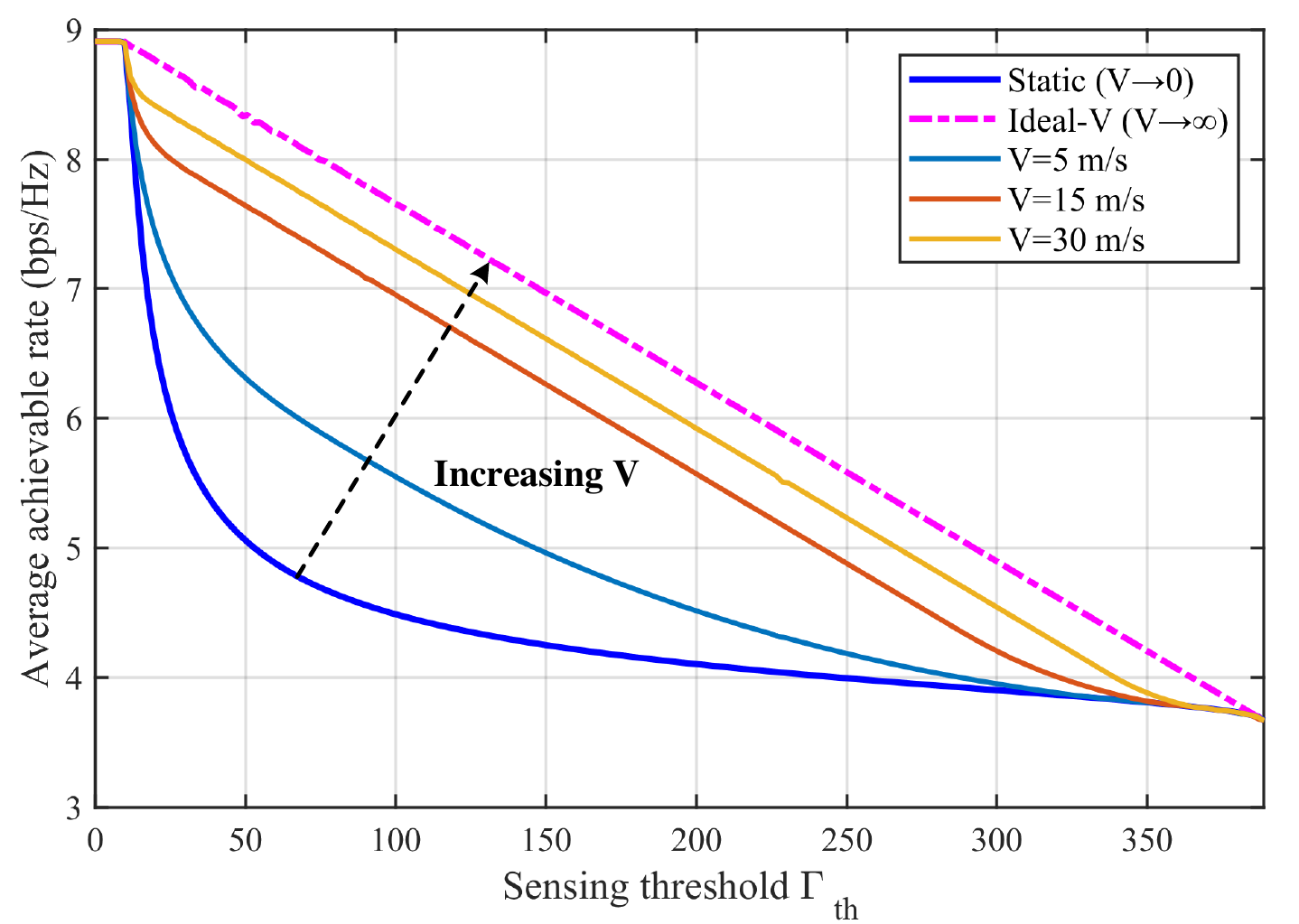}
\caption{Achievable rate versus sensing threshold $\Gamma_{\mathrm{th}}$ under dynamic trajectories.}
\label{fig_4}
\vspace{-15pt} 
\end{figure}

Fig.~4 plots the achievable rate versus the sensing SNR threshold under the values of different UAV speeds. The parameters are the same as those for Fig.~2. The results clearly demonstrate that increasing UAV speed enlarges the achievable rate region and rapidly approaches the ideal upper bound. This highlights the pronounced benefits of dynamic trajectory design, particularly under stringent sensing requirements.
Moreover, the rate loss induced by increasing sensing demands is considerably alleviated for higher UAV speeds, illustrating the critical role of UAV mobility in achieving robust ISAC performance across a broad range of system constraints.

\section{Conclusion}

This paper proposed a dual-level channel reconfiguration framework for RA-enabled UAV ISAC that jointly exploits the UAV mobility and the RA rotation to proactively reconfigure the path loss and the subspace correlation of S\&C channels. For the scenario of static UAV deployment, we derived the closed-form expressions of the optimal RA rotation, beamforming, and hovering location, yielding a more flexible trade-off between the performance of S\&C. Both the analytical results and numerical results demonstrated the significant gains from RA rotation and position optimization, and also revealed additional performance improvement via carrying out time-sharing, highlighting the potential of UAV mobility. Inspired by this result, we then investigated a fully mobile UAV and unveiled the optimality of the HFH UAV trajectory, thereby obtaining the globally optimal solution. Simulations validated our analytical results and also demonstrated the superiority of the proposed scheme over various benchmark schemes.

\section*{Appendix A: \textsc{Proof of Theorem 2}}
The feasible set of problem (14) is 
\begin{equation}
\mathcal S = \{\mathbf q\in\mathbb R^2:\ \|\mathbf q-\mathbf q_t\|\le R_h\},
\quad R_h=\sqrt{R_s^2-H^2}.
\end{equation}
Since \(R(\mathbf q)=\log_2\!\big(1+\frac{A}{\|\mathbf q-\mathbf q_u\|^2+H^2}\big)\) is strictly
decreasing in \(\|\mathbf q-\mathbf q_u\|\), problem~\eqref{eq:P3} is equivalent to
\begin{equation}\label{eq:projProb}
\min_{\mathbf q\in \mathcal S}\ \|\mathbf q-\mathbf q_u\|.
\end{equation}
If $\Gamma_{\mathrm{th}}\le \Gamma(\mathbf q_u)=B/(D_h^2+H^2)$, then constraint (14b) is inactive, and the optimal deployment is to place the UAV right above the user with $\mathbf q^*=\mathbf q_u$, yielding $R^*(\Gamma_{\mathrm{th}})=\log_2\!\big(1+\dfrac{A}{H^2}\big)$. Otherwise, when $\dfrac{B}{D_h^2+H^2}<\Gamma_{\mathrm{th}}\le\dfrac{B}{H^2}$, (14b) is active, and the optimal point must satisfy $\|\mathbf q-\mathbf q_t\|^2+H^2=\dfrac{B}{\Gamma_{\mathrm{th}}}$. Since $R(\mathbf q)$ strictly decreases with respect to $\|\mathbf q-\mathbf q_u\|$, the optimal $\mathbf q^*$ lies on the line segment between $\mathbf q_t$ and $\mathbf q_u$, satisfying $\|\mathbf q^*-\mathbf q_t\|=R_h$. Thus, $\mathbf q^*=\mathbf q_t+\dfrac{R_h}{D_h}(\mathbf q_u-\mathbf q_t)$. In this case, $\|\mathbf q^*-\mathbf q_u\|=D_h-R_h$, and thus we have (17).\hfill$\blacksquare$

\section*{Appendix B: \textsc{Proof of Proposition 1}}
For the static deployment, the closed-form Pareto boundary ${{\cal C}_{{\mathop{\rm sta}\nolimits} }}$ is expressed in (18). For notational simplicity, we henceforth write $\Gamma \equiv \Gamma_{\mathrm{th}}$. For $\Gamma\in\big(\frac{B}{D_h^2+H^2},\,\frac{B}{H^2}\big)$, taking the second derivative of $R^*(\Gamma_{\mathrm{th}})$ with respect to $\Gamma$ yields
\begin{equation}
    R''(\Gamma) = -\frac{A}{\ln 2} \frac{M(\Gamma)}{[g(\Gamma)\,(g(\Gamma)+A)]^2},
\end{equation}
where $s(\Gamma) = \sqrt{\frac{B}{\Gamma} - H^2}$,  $g(\Gamma) = (D_h - s)^2 + H^2$, and $M(\Gamma) = B^2 g(g+A)D_h - 2B^2 (D_h-s)^2 s (2g+A) - 4B (D_h-s) \Gamma s^2 g(g+A)$,
with the domain $\Gamma \in (\Gamma_{\min}, \Gamma_{\max})$, $\Gamma_{\min} = \frac{B}{D_h^2 + H^2}$, and $\Gamma_{\max} = \frac{B}{H^2}$.
It is observed that, at both endpoints, $R''(\Gamma) < 0$.
For certain values of $D_h$ and $H$, there exists $\Gamma$ in the domain for which $R''(\Gamma)>0$. Specifically, $R''(\Gamma) > 0$ if
\begin{equation}
    D_h^2 > \frac{4}{3} \sqrt{H^2 + A} \left( \sqrt{4H^2 + A} - \sqrt{H^2 + A} \right),
\end{equation}
which implies that ${{\cal C}_{{\mathop{\rm sta}\nolimits} }}$ is non-convex.
\hfill$\blacksquare$

\vfill

\end{document}